\documentclass[a4paper]{jpconf}

\usepackage{amsmath}
\usepackage{amssymb}
\usepackage{graphicx}
\usepackage{epsf}
\usepackage{slashed}
\usepackage{enumitem}
\usepackage[czech,english]{babel}
\usepackage[cp1250]{inputenc}
\usepackage{hyperref}
\usepackage{xcolor}

\usepackage[only,llbracket,rrbracket,llparenthesis,rrparenthesis]{stmaryrd} 
\usepackage{accsupp} 

\newcommand{\I}{\ensuremath{\mathrm{i}}}
\newcommand{\eL}{\mathcal{L}}
\renewcommand{\d}{\ensuremath{\mathrm{d}}}
\newcommand{\qm}[1]{``#1''} 

\usepackage{bbm} 
\newcommand{\unitmatrix}[1]{\ensuremath{\mathbbm{1}_{#1}}}

\begin{document}

\title{Cho--Maison magnetic monopole: BPS limit and lower mass bound}

\author{Petr Bene\v{s}$^1$ and Filip Blaschke$^{1,2}$}
\address{$^1$ Institute of Experimental and Applied Physics,\\Czech Technical University in Prague,\\Husova 240/5, 110~00 Prague 1, Czech Republic}
\address{$^2$ Institute of Physics and Research Centre of Theoretical Physics and Astrophysics,\\Faculty of Philosophy and Science, Silesian University in Opava,\\Bezru\v{c}ovo n\'am\v{e}st\'i~1150/13, 746~01 Opava, Czech Republic}


\ead{petr.benes@utef.cvut.cz}

\begin{abstract}
We construct a class of effective extensions of the Standard Model that support finite-mass Cho--Maison magnetic monopole in the Bogomol'nyi--Prasad--Sommerfield (BPS) limit. We present an example of exact analytic monopole solution and derive a universal lower mass bound $M \geq 2\pi v/ g \approx 2.37\,\,\mathrm{TeV}$ of the Cho--Maison magnetic monopole.
\end{abstract}

\section{Introduction}

It's been a well known fact that in the Standard Model (SM) there is no magnetic monopole of the 't~Hooft--Polyakov type \cite{tHooft:1974kcl,Polyakov:1974ek}. Reason is that the second homotopy group of the vacuum manifold is trivial, $\pi_2\big(SU(2)_{\mathrm{L}}\times U(1)_{\mathrm{Y}}/U(1)_{\mathrm{em}}\big) = \{1\}$. However, it was shown by Cho and Maison \cite{Cho:1996qd} that the desired topology that could support the solitonic monopole solution can be found elsewhere, namely in the Higgs doublet field $H$ itself. In order to see that, it is instrumental to write $H$ as
\begin{eqnarray}
H &=& \frac{v}{\sqrt{2}}\,\rho\,\xi \,,
\end{eqnarray}
where $\xi$ is a complex doublet normalized as $\xi^\dag \xi = 1$ and $v = 246\,\mathrm{GeV}$ is the vacuum expectation value of $H$ (the Higgs potential is $\propto (|H|^2-v^2/2)^2$). It can be shown that, due to the $U(1)_{\mathrm{Y}}$ symmetry, the field $\xi$ can be regarded as a $\mathbb{C}\mathrm{P}^1$ coordinate. But $\pi_2(\mathbb{C}\mathrm{P}^1) = \mathbb{Z}$, so that stable soliton solutions with the properties of a magnetic monopole can, potentially, exist in SM.

Indeed, such a solution can be found. Alas, it turns out that it has an infinite mass:
\begin{equation}
M \ = \ \frac{2\pi}{g^{\prime\,2}} \int_{0}^{\infty} \frac{\d r}{r^2} + \mbox{finite terms} \ = \ \infty \,.
\end{equation}
Thus, the common wisdom that there is no (physical, finite-mass) magnetic monopole in SM seems justified.

Thus, one must step beyond the SM and modify it in order to get a finite-mass monopole. One way to do that is due to Cho, Kim, and Yoon (CKY) \cite{Cho:2013vba} and consists of modifying the hypercharge kinetic term as
\begin{equation}
-\frac{1}{4g^{\prime\,2}} B_{\mu\nu} B^{\mu\nu}
\hspace{3mm} \longrightarrow \hspace{3mm}
-\,\frac{1}{4g^{\prime\,2}} \epsilon\big( |H|^2 \big) B_{\mu\nu} B^{\mu\nu} \,,
\end{equation}
where $\epsilon$ is some positive function of the Higgs field squared, normalized as $\epsilon(v^2/2) = 1$ (in order to recover the SM in the vacuum). Crucially, if $\epsilon \to 0$ sufficiently fast as $|H|^2 \to 0$, the mass of the Cho--Maison monopole comes out finite, as desired.
In fact, models beyond SM of this type have indeed been proposed in the literature \cite{Arai:2017lfv,Arai:2017ntb}.

Since $\epsilon$ is otherwise virtually arbitrary, also the corresponding monopole mass is in principle arbitrary. E.g., Ellis, Mavromatos and You \cite{Ellis:2016glu} were varying $\epsilon$ (while keeping consistency with experimental bounds on $H \to \gamma\gamma$ production) and found a whole variety of monopole masses reaching as low as to $5.5\,\,\mathrm{TeV}$. Thus, a question of phenomenological importance arises whether one can go with monopole mass arbitrarily low or if there exists some limit.

Indeed, as showed recently, it turns out that there exists a definite lower bound for the monopole mass \cite{Blaschke:2017pym}. In order to derive it, one has to first construct a Bogomol'nyi--Prasad--Sommerfield (BPS) limit of the Cho--Maison monopole and show that such the mass bound holds also in the CKY theory.

\section{Construction of the BPS limit}

In order to construct a BPS limit of a given theory, it is usually sufficient to express the Hamiltonian of a static configuration as a perfect square plus a total derivative and nullify what remains (typically the potential). For the Cho--Maison monopole (with CKY regularization) this approach doesn't work, though. It turns out that besides removing the potential one has to modify non-trivially the gauge sector of the theory, too. Instead, we turn the typical procedure of constructing the BPS theory upside down and start from the desired result: We first write down the BPS equation of motion and subsequently derive the BPS theory from it.

Thus, we postulate the most general BPS equation
\begin{eqnarray}
\label{BPSeqgeneral}
D_i H &=& \frac{1}{\rho} f_1(\rho)\, M_i \xi + f_2(\rho)\, \big(\xi^\dag M_i\,\xi\big)\xi + f_3(\rho)\, G_i \xi \,,
\end{eqnarray}
where we defined \qm{magnetic fields} of $SU(2)_\mathrm{L}$ and $U(1)_\mathrm{Y}$
\begin{equation}
M_i \ = \ \frac{1}{2} \varepsilon_{ijk}F_{jk}^a \sigma_a \,,
\hspace{1cm}
G_i \ = \ \frac{1}{2} \varepsilon_{ijk}B_{jk} \,.
\end{equation}
Notice presence of as yet arbitrary functions $f_1$, $f_2$, $f_3$ of $\rho = \sqrt{2} |H|/v$.

The equation \eqref{BPSeqgeneral} is gauge-covariant, linear in the gauge fields and first-order in derivatives. Thus, it qualifies as a candidate for a valid BPS equation, that is, equation of motion for a BPS theory. This operationally means that Hamiltonian of a static configuration of such a theory should be possible to write in the form
\begin{eqnarray}
\mathcal{H} &=& \bigg|D_i H-\frac{1}{\rho} f_1\, M_i \xi - f_2\, \big(\xi^\dag M_i\,\xi\big)\xi - f_3\, G_i \xi \bigg|^2
+ \partial_i X_i
\end{eqnarray}
for some $X_i$, that is, as a sum of squares that vanish when the BPS equation is satisfied, plus a total derivative. Moreover, one also requires that the Hamiltonian be non-negative. This can indeed be done, but only if the functions $f_i$ are no longer independent, but $f_2$ is related to $f_1$ as
\begin{eqnarray}
f_2 &=& \frac{1}{2} f_1^\prime - \frac{1}{\rho} f_1
\end{eqnarray}
(prime denotes derivative with respect to $\rho$). The three vector $X_i$ is then given as
\begin{eqnarray}
X_i &=& \frac{v}{\sqrt{2}} f_1\, \big(\xi^\dag M_i\,\xi\big) + \sqrt{2} v F_3 \, G_i \,,
\hspace{1cm}\mbox{where}\hspace{0.5cm}
F_3^\prime = f_3 \,,
\end{eqnarray}
and the Hamiltonian can be written in a manifestly non-negative form
\begin{eqnarray}
\mathcal{H} &=& 
\big|D_i H\big|^2 + \bigg|\frac{1}{\rho} f_1 \Big(\unitmatrix{} -\xi \xi^\dag\Big) M_i \xi + \frac{1}{2} f_1^\prime \big(\xi^\dag M_i\,\xi\big)\xi + f_3\, G_i \xi \bigg|^2 \,.
\end{eqnarray}

Now, when we know the BPS Hamiltonian, we can write down the corresponding Lagrangian:
\begin{eqnarray}
\label{LagrangianBPS}
\eL_{\mathrm{BPS}}
&=& |D_\mu H|^2 
-\frac{v^2}{4 g^2 |H|^2} h^2 \bigg( \Tr\big[F_{\mu\nu}^2\big] - \frac{2}{|H|^4} \Tr\big[F_{\mu\nu} H H^{\dagger}\big]^2
 \bigg)
\nonumber \\ && \phantom{|D_\mu H|^2}
-\frac{1}{4} \bigg(\frac{h^\prime}{g|H|^2} \Tr\big[F_{\mu\nu} H H^{\dagger}\big] + \frac{f^\prime}{g^\prime} B_{\mu\nu}\bigg)^2 \,,
\end{eqnarray}
where we rescaled for convenience the functions $f_1$ and $f_3$ in favor of functions $f^\prime$ and $h$ as
\begin{equation}
f_1 = \frac{\sqrt{2}}{g} h \,,
\hspace{10mm}
f_3 = \frac{1}{\sqrt{2}g^\prime} f^\prime \,.
\end{equation}
We assume $f^\prime(1) = h(1)  = 1$ in order to have standard normalization of kinetic term in vacuum ($\rho = 1$). The BPS equation reads
\begin{eqnarray}
D_i H &=&
\frac{\sqrt{2}}{g\rho} h(\rho)\, \Big(M_i -\xi^{\dagger} M_i  \xi\Big)\xi
+ \frac{1}{\sqrt{2}g} h^\prime(\rho)\, \Big(\xi^{\dagger} M_i  \xi\Big)\xi
 +\frac{1}{\sqrt{2}g^{\prime}}f^{\prime}(\rho)\,G_i\, \xi \,.
\end{eqnarray}
A consequence of this equation is that the $U(1)_\mathrm{Y}$ gauge field $B_i$ has the form
\begin{eqnarray}
\label{Bi}
B_i &=& 2 \I \xi^\dag \bigg(\partial_i + \frac{1}{2}\I A_i^a \sigma^a \bigg) \xi \,,
\end{eqnarray}
i.e., it is entirely expressed in terms of the $SU(2)_\mathrm{L}$ gauge field $A_i^a$ and $\xi$.

If the BPS equation is satisfied, the Hamiltonian is given only by the surface term, i.e.,
\begin{eqnarray}
\label{HamiltonianBPS}
{\mathcal H} &=& \partial_i \bigg[
\frac{v}{g} h(\rho)\, \xi^{\dagger} M_i \xi  + \frac{v}{g^{\prime}}f(\rho)\,G_i \bigg] \,,
\end{eqnarray}
i.e., the Bogomol'nyi bound is saturated.

\section{Exact monopole solution - An example}

The BPS equation is first order, not second order as a usual equation of motion, so there is a hope that it might be exactly solvable, at least for suitable $f^\prime$ and $h$. Indeed, several exact solutions have been found \cite{Blaschke:2017pym}.

Adopting the spherically symmetric Ansatz for the $SU(2)_\mathrm{L}$ gauge field
\begin{subequations}
\label{ansatz}
\begin{eqnarray}
A_i^a &=& \big(1-K\big)\varepsilon_{iak}\frac{x_k}{r^2} \,,
\end{eqnarray}
where $K = K(r)$ is some function of the radial coordinate, and assuming $\xi$ to have the form (in spherical coordinates)
\begin{eqnarray}
\xi &=& \I
\left(\begin{array}{c} \sin(\theta/2) \, \e^{-\I\varphi} \\ -\cos(\theta/2) \end{array}\right)
\end{eqnarray}
\end{subequations}
(so that $\xi$ is a member of the lowest non-trivial homotopy class), the problem is reduced to finding the two unknown functions of the radial coordinate, $K$ and $\rho$, subject to the boundary conditions
\begin{subequations}
\label{boundarycondt}
\begin{align}
K(0)     &=1 \,, & \rho(0)     &=0 \,,  \\
K(\infty)&=0 \,, & \rho(\infty)&=1 \,, 
\end{align}
\end{subequations}
dictated by requirements of regularity of the solution in the origin and of finiteness of the total energy. It is not hard to show that with this Anstaz the equation \eqref{Bi} for $B_i$ yields
\begin{equation}
B_i = 2i \xi^{\dagger}\partial_i \xi = (1-\cos\theta) \partial_i\varphi \ = \ -\varepsilon_{ij3} \partial_j \log(r+z) \,,
\end{equation}
regardless of the form of $K$.

Here we present, for the sake of an illustration, only the simplest solution, corresponding to the BPS theory defined by functions
\begin{equation*}
f^\prime(\rho) \ = \  \rho^2 \ = \ \frac{2 |H|^2}{v^2} \,,
\hspace{10mm}
h(\rho) \ = \ 1 \,.
\end{equation*}
The resulting $\rho$ and $K$ read
\begin{subequations}
\begin{eqnarray}
\rho(r) &=& \frac{1}{\displaystyle 1+\frac{1}{\mu r}}
\,,
\\
K(r) &=& \exp\bigg\{\!\! -\frac{g}{g^\prime} \bigg[ \frac{\mu r}{2} \, \frac{2+\mu r}{1+\mu r} - \log\big(1+\mu r\big)  \bigg] \bigg\} \,,
\end{eqnarray}
\end{subequations}
where we defined for convenience the scale $\mu \equiv v g^{\prime} \approx 86.1 \,\,\mathrm{GeV}$. This solution corresponds to a monopole with the magnetic charge
\begin{eqnarray}
q &=& \frac{4\pi}{e} \,,
\end{eqnarray}
where $e = gg^\prime / \sqrt{g^2+g^{\prime2}}$ is the electric charge, and mass
\begin{eqnarray}
M &=& 4\pi v \bigg(\frac{1}{2g} + \frac{1}{3 g^\prime}\bigg)  \ \approx \ 5.32\,\,\mathrm{TeV} \,.
\end{eqnarray}

\section{Monopole mass and its lower bound}

Recall that energy density $\mathcal{H}$ is given by expression \eqref{HamiltonianBPS}. Since it is a total derivative, the total energy (mass) of the system $M = \int_{\mathbb{R}^3} \d^3 x \, \mathcal{H}$ doesn't depend, due to the Gauss--Ostrogradsky theorem, on details of $A_i^a$, $B_i$ and $H$ in the whole space volume, but only on their asymptotic behavior.

For our spherically symmetric Ansatz \eqref{ansatz} this asymptotic behavior is given solely by the boundary conditions \eqref{boundarycondt}. Thus, it is possible, without having to solve the equations of motion at all, to arrive at a simple formula for the magnetic monopole mass:
\begin{eqnarray}
M &=& 
4\pi v \bigg[ \frac{1}{2g} + \frac{1}{g^\prime} \, \Big| \int_0^1\!\d \rho \, f^\prime(\rho) \Big| \bigg]
\,.
\end{eqnarray}
(Curiously, this expression depends only on $f^\prime$, not on $h$.) From this result it immediately follows that there is a lower bound on the mass of the monopole:
\begin{eqnarray}
\label{massbound}
M &\geq&  \frac{2\pi v}{g} \ \approx \ 2.37\,\,\mathrm{TeV} \,.
\end{eqnarray}

Thus, in a BPS theory \eqref{LagrangianBPS} the mass of a magnetic monopole cannot be lower than $2\pi v/g$, regardless of the choice of functions $f^\prime$ and $h$. However, the BPS theory doesn't describe the real world. The question therefore arises whether this bound holds also in a more realistic non-BPS theory, that is, in particular, in the SM with CKY regularization \cite{Cho:2013vba}, whose bosonic part reads
\begin{eqnarray}
\eL_{\mathrm{CKY}} &=& |D_\mu H|^2 - V\big(|H|^2\big) -\frac{1}{4 g^2} \Tr\big[F_{\mu\nu}^2\big] -\frac{1}{4 g^{\prime 2}} \epsilon(\rho) B_{\mu\nu}^2 \,,
\end{eqnarray}
where, again, $\rho = \sqrt{2} |H| / v$.

Usually (e.g., for the 't~Hooft--Polyakov monopole) the only difference between a BPS and non-BPS theory is lack or presence of the scalar potential. Since it is positive, the energy density (Hamiltonian) of a BPS theory is never greater than that of a non-BPS theory. Thus, any lower mass bound holding in the former holds necessarily also in the latter.

In the present case the situation is more complicated. One might wish that
\begin{eqnarray}
\label{inequality}
\mathcal{H}_{\mathrm{BPS}}[f^\prime,h] &\leq& \mathcal{H}_{\mathrm{CKY}}[\epsilon]
\end{eqnarray}
for \emph{arbitrary} $f^\prime$, $h$ and $\epsilon$. This, however, turns out not to be the case. Nevertheless, this requirement is in fact too strong anyway. It suffices to demand that for each $\epsilon$ there exist $f^\prime$ and $h$ such that the inequality \eqref{inequality} holds. This weaker requirement already proves to be true. E.g., for $f^\prime$, $h$ defined in terms of $\epsilon$ as
\begin{equation}
f^\prime(\rho) \ = \ \sqrt{\frac{3}{4}\epsilon(\rho)} \,,
\hspace{10mm}
h(\rho) \ = \ \rho 
\end{equation}
the inequality \eqref{inequality} holds indeed. This proves that the magnetic monopole bound \eqref{massbound} holds also in a more realistic non-BPS theory.

\section{Conclusions}

In this note, we have presented a family of effective extensions of the SM that contain BPS magnetic monopoles as classical solutions. These solutions are the BPS extensions of Cho--Maison monopoles presented in \cite{Cho:2013vba} and further studied in \cite{Ellis:2016glu}. We have also presented an explicit example of monopole solution in analytic form. 

Critically, we obtained a universal lower bound on the mass of the monopole as $M \geq 2\pi v/ g \approx 2.37\,\,\mathrm{TeV}$ and argued that this bound also applies to phenomenologically relevant non-BPS models, in particular to the CKY model presented in \cite{Cho:2013vba} and \cite{Ellis:2016glu}.

Our work opens up the possibility of studying multi-particle configurations of Cho--Maison monopoles, which are sometimes characterized as conceptually being something between Dirac's monopole and the 't~Hooft--Polyakov monopole. To construct a multi-monopole configuration of Dirac's monopoles is not particularly challenging. On the other hand, to obtain a 't~Hooft--Polyakov multi-monopole configuration requires the use of highly sophisticated tools, such as the Nahm construction \cite{Nahm:1981nb}. It would make an interesting future study to elaborate how difficult is to write down a multi-monopole configuration of Cho--Maison monopoles in the BPS limit and whether one can adopt the Nahm approach or other techniques.

Furthermore, our model can be a useful tool to explore non-topological solutions of the electroweak model, such as the spharelon \cite{Klinkhamer:1984di}. Indeed, the BPS limit lends itself to the possibility that we can find unstable static solutions in the analytic form. We plan to investigate this in the future work.

As far as we know our theory is not a bosonic part of any known supersymmetric theory. It would be interesting to see if there actually exists a SUSY extension of our model and whether the BPS Cho--Maison monopole can be constructed as a 1/2, 1/4, or other SUSY state. We leave this puzzle for the future.

\ack

This work was supported by the Albert Einstein Centre for Gravitation and Astrophysics financed by the Czech Science Agency Grant No.~14-37086G (F.~B.) and by the program of Czech Ministry of Education, Youth and Sports INTEREXCELLENCE Grant No.~LTT17018 (F.~B., P.~B.). One of the authors, P.~B., thanks TJ~Balvan Praha for support and the organizers of the ISQS-26 for the opportunity to speak at the conference.

\section*{References}


\providecommand{\newblock}{}

\end{document}